\documentclass[twoside,slac_one]{revtex4}
\usepackage{graphicx}
\usepackage{fancyhdr}
\usepackage{amsmath} 
\usepackage{bm}
\usepackage{amsxtra}
\usepackage{amssymb}
\usepackage{amsthm}
\usepackage{latexsym}
\usepackage{lscape}
\usepackage{textcomp}

\begin{document}

\title{NO$\nu$A: Present and Future}

%

\author{Gavin S. Davies for the NO$\nu$A Collaboration}
\affiliation{Department of Physics and Astronomy, Iowa State University, Iowa, IA, USA}

\begin{abstract}
NO$\nu$A is a next generation neutrino oscillation experiment designed to search
for muon neutrino to electron neutrino oscillations by comparing electron neutrino event rates in a Near Detector at Fermilab with the rates observed in a large Far Detector at Ash River, Minnesota, 810~km from Fermilab. The detectors are totally active, segmented, liquid scintillator detectors and are located 14~mrad off the NuMI beam axis. The Far Detector has begun construction and will begin taking data in early 2013. The experiment aims to measure the neutrino mixing angle $\theta_{13}$ and will push the search for electron neutrino appearance beyond the current limits by more than an order of magnitude. For non-zero $\theta_{13}$, it is possible for NO$\nu$A to observe CP violation in neutrinos and establish the neutrino mass hierarchy. The NO$\nu$A prototype near detector on
the surface (NDOS) began running at Fermilab in November and registered its first neutrinos from the NuMI beam in December 2010. An overview and current status of the experiment will be presented.
\end{abstract}

\maketitle

\thispagestyle{fancy}


\section{Introduction}
The NuMI\footnote{Neutrinos at the Main Injector} Off-Axis $\nu_{e}$ Appearance experiment (NO$\nu$A) \cite{nova,nova2} is the flagship project of Fermilab’s Intensity Frontier initiative and will study the appearance of electron neutrinos or antineutrinos using the NuMI neutrino beam produced at Fermilab. NO$\nu$A is a next generation neutrino oscillation experiment with a baseline of 810~km. The beam is measured by a Near Detector (ND) about 1~km downstream of the beam production target and a Far Detector (FD) placed 810~km away in Ash River, Minnesota. The detectors are positioned 14~mrad (0.8\textdegree) off the NuMI beam axis so that the expected neutrino beam energy spectrum is a narrow band peak at 2~GeV near the oscillation maximum. This yields a dramatic reduction of backgrounds in $\nu_{e}$ appearance searches.

\section{The NO$\nu$A Experiment}

\subsection{NuMI Beam}
\begin{figure}[ht]
\begin{minipage}[t]{0.5\linewidth}
\centering
\includegraphics[width=80mm]{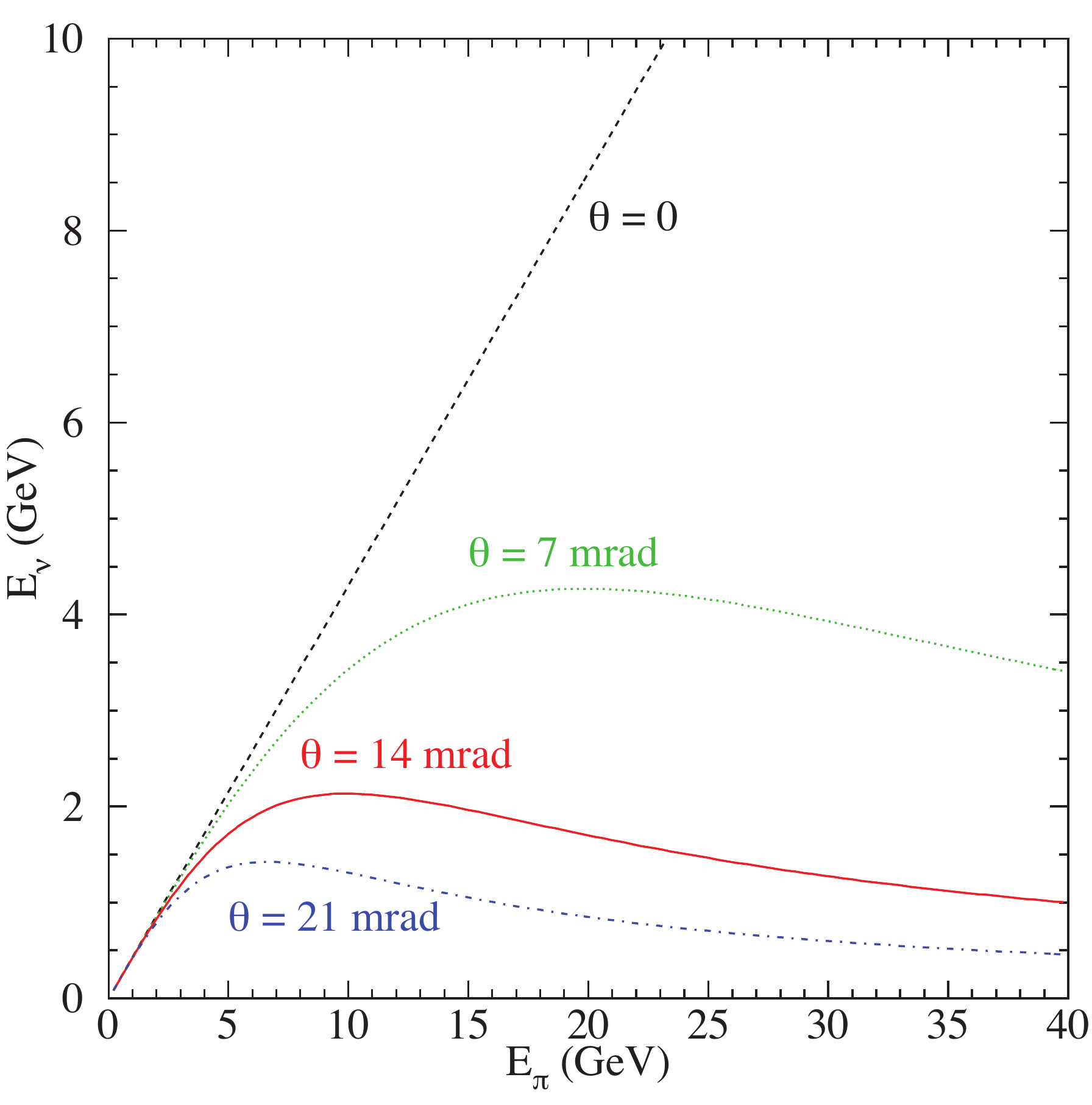}
\caption{Neutrino energies as a function of the energy of the parent pion and the pion decay angle $\theta$.} \label{fig:neutrinoE}
\end{minipage}%
\begin{minipage}[t]{0.5\linewidth}
\centering
\includegraphics[width=80mm]{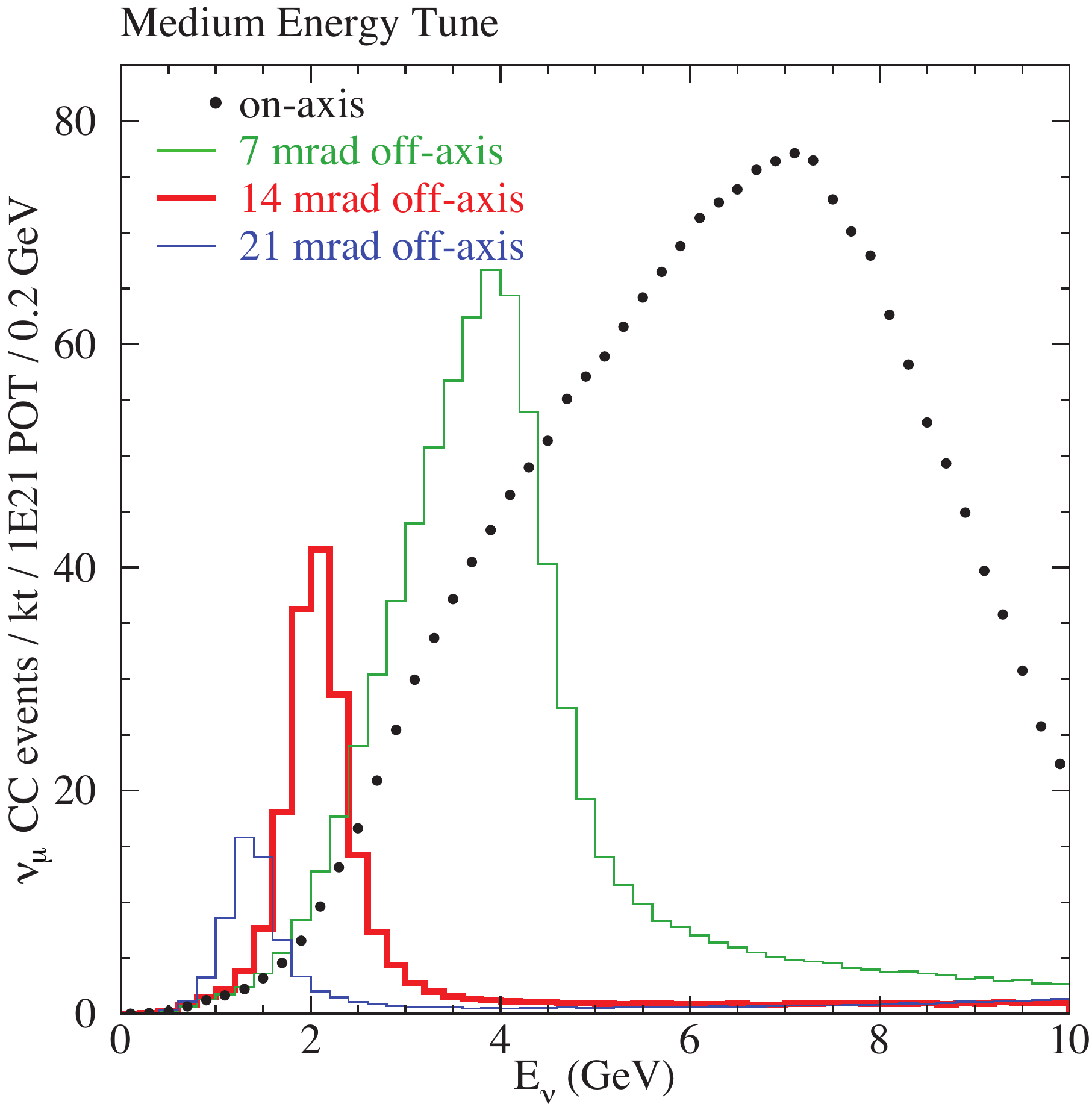}
\caption{the expected neutrino beam energy spectra for different off-axis detector locations in a medium energy tune in NuMI compared to the on-axis energy spectrum.} \label{fig:neutrinoE2}
\end{minipage}
\end{figure}

NO$\nu$A will study $\nu_{\mu} \rightarrow \nu_{e}$ oscillations in the NuMI neutrino beam. The beam is created from collisions of 120~GeV protons, accelerated by the Main Injector facility at Fermilab, with a graphite target. To meet the NO$\nu$A physics goals, the NuMI beam will be upgraded from a nominal power of 400~kW to 700~kW of beam power. This will be achieved by reducing the cycle time of the Main Injector from the present 2.2~s to 1.3~s via slip-stacking in the recycler ring; increasing the intensity per cycle with 12 Booster batches instead of 11 by installing new RF stations and a new injection kicker magnet; and upgrading the target and horns to accommodate the
increased proton intensity. There will be a 10~$\mu$s beam spill every 1.33~s as a result. The upgrades will be carried out during the scheduled March 2012 shutdown of the Fermilab accelerator facilities and will result in an intensity of 4.9 x 10$^{13}$ protons per pulse, corresponding to a total 6.0 x 10$^{20}$ protons on target (POT) accumulated per year of running. From pion decay kinematics, we know the neutrino energy will depend on the characteristic decay angle $\theta$ between the neutrino and the parent pion in the lab frame. As shown in Fig.~\ref{fig:neutrinoE}, for $\theta$ = 14~mrad most pion decays result in neutrinos with E = 2~GeV, with some energy smearing around that value. Therefore, the NO$\nu$A detectors placed 14~mrad off the NuMI beam axis will measure a narrow band beam peaked very near the $\nu_{\mu} \rightarrow \nu_{e}$ oscillation maximum at 810~km (E $\sim$1.6 GeV), as shown in Fig.~\ref{fig:neutrinoE2}. The narrow beam energy spectrum strongly reduces the background from feed-down of higher energy neutral-current (NC) neutrino events, the dominant background for $\nu_{\mu} \rightarrow \nu_{e}$ oscillation searches with the detectors placed on-axis.

\subsection{Near and Far Detectors}

The NO$\nu$A detector system consists of a near and far detector constructed 14~mrad off-axis
to the NuMI source. The detectors are functionally identical, allowing for cancellation of systematic uncertainties due to neutrino flux and cross-section modeling. Both detectors are highly segmented tracking calorimeters built entirely from low Z ($\sim$0.15 radiation lengths per layer), highly reflective (15\% TiO2) PVC cells with a 65\% active volume \cite{nova}. The PVC cells are filled with liquid scintillator (mineral oil infused with 5\% pseudocumene). Each cell is 3.8~cm by 5.9~cm in cross section with 90\% reflectivity for light at 430~nm. Sixteen cells are extruded together as a single part. Two extrusions are joined together by an end seal and manifold cover to produce a sealed module of 32 cells. The orientation of the cells alternates between vertical and horizontal to allow for 3D event reconstruction. The scintillation light is collected by a loop of wavelength shifting fiber in each cell and read out by 32-pixel avalanche photo-diodes (APDs). The APDs have 85\% quantum efficiency, operate at a gain of 100 and are cooled to -15\textdegree~C to reduce dark noise to less than 2 photoelectrons.

The detectors are optimized for detection of $\nu_{e}$ charged-current (CC) interactions, with fine sampling of the characteristic electromagnetic showers (1 layer = 0.15 X$_{0}$, Moli\`{e}re radius = 10~cm). The ND will have a mass of 222~ton and will be placed underground, in a new cavern to be excavated next to the MINOS \cite{minos} experiment ND hall, $\sim$ 1~km away from the beam production target. Excavation of the underground cavern for the near detector will begin following the beam shutdown. The ND will observe a cosmic rate of 50~Hz and will see approximately 30 neutrino events per beam spill. 

\begin{figure}[ht]
\begin{minipage}[t]{0.5\linewidth}
\centering
\includegraphics[width=80mm]{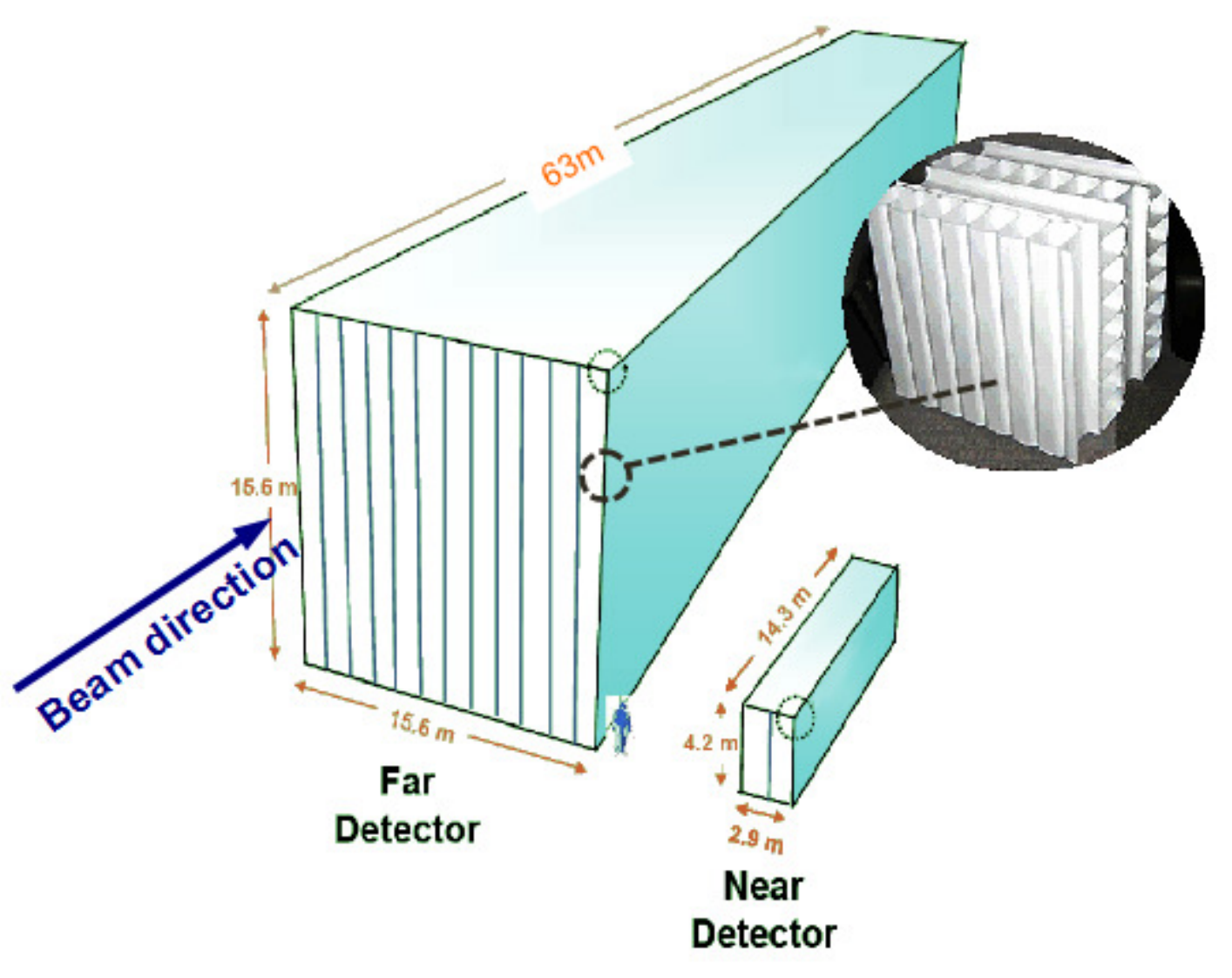}
\caption{The relative sizes of the NO$\nu$A experiment's Near and Far  detectors.} \label{fig:relsize}
\end{minipage}%
\begin{minipage}[t]{0.5\linewidth}
\centering
\includegraphics[width=80mm]{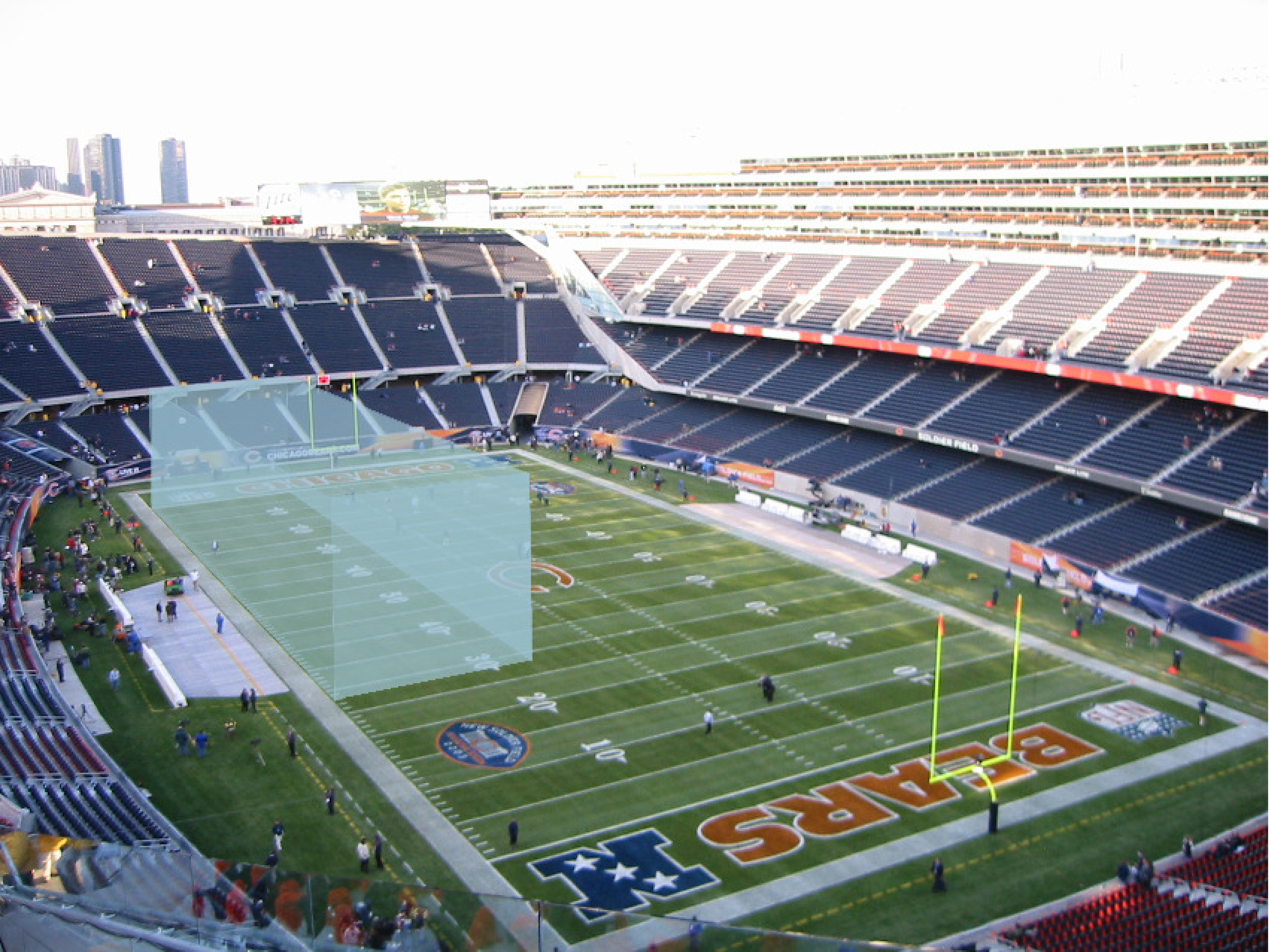}
\caption{The Far Detector superimposed in Soldier Field stadium.} \label{fig:soldierField}
\end{minipage}
\end{figure}

The much larger FD will have a mass of 14~kton and will be located at the surface in Ash River, Minnesota, 810~km away.
The relative size of the two detectors is apparent in Fig.~\ref{fig:relsize} and the scale of the experiment is highlighted by superimposing the Far Detector in Chicago's Soldier Field as in Fig.~\ref{fig:soldierField}. The FD will have a cosmic rate  of approximately 200~kHz and will see on the order of 50 $\nu_{e}$ charged-current signal events per year.

\begin{figure}[ht]
\centering
\includegraphics[width=135mm]{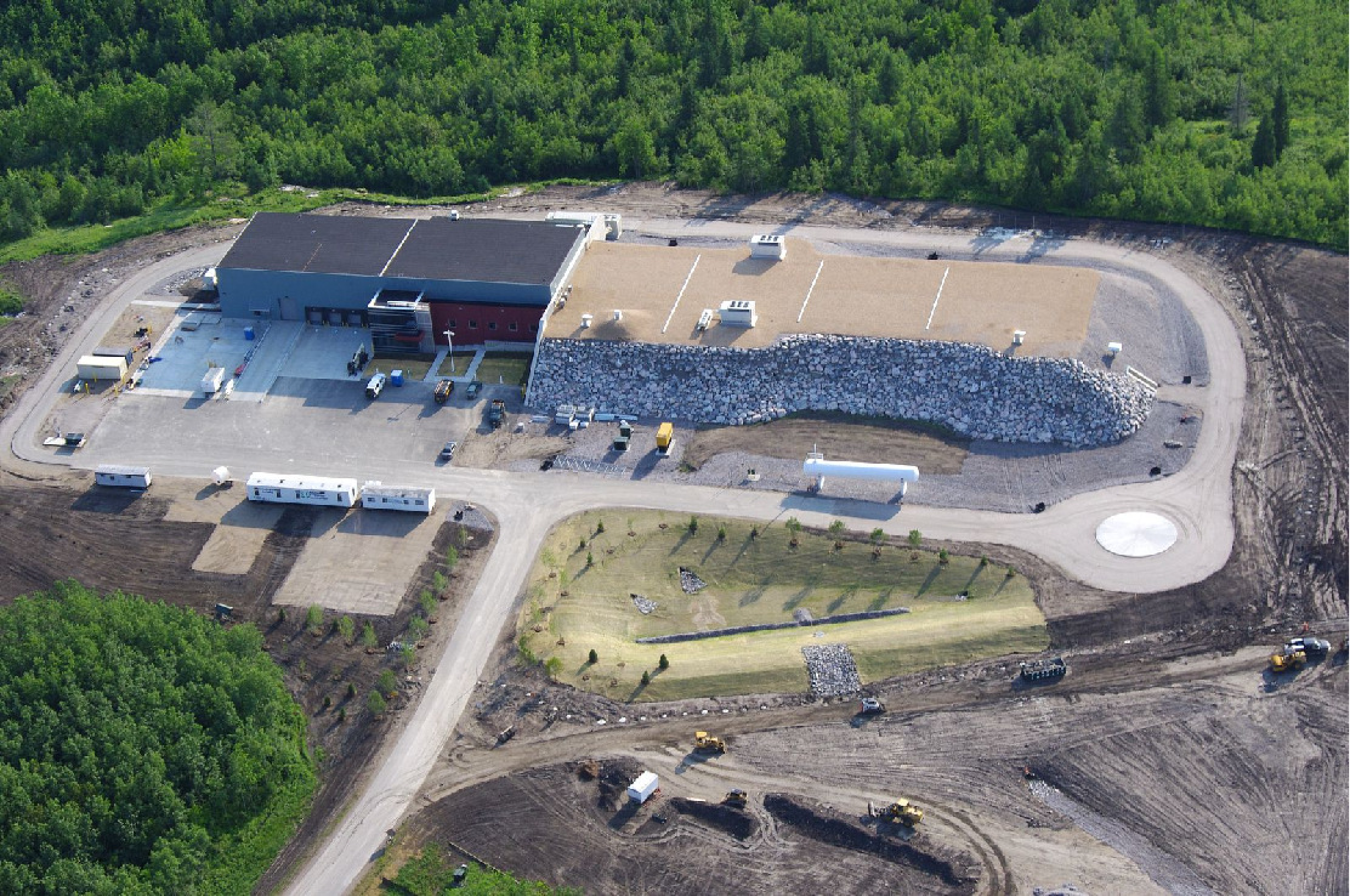}
\caption{Aerial view photograph of the NO$\nu$A Far Detector site in Ash River, Minnesota.} \label{fig:fardetView}
\end{figure}

The FD building at Ash River includes a mound of barite rock 3~m earth-equivalent overburden of more than ten radiation lengths to reduce cosmic backgrounds. The FD PVC blocks will measure 15.6~m in height by 15.6~m in width. The total length of the detector will exceed 63~m. The FD PVC modules will be assembled and tested in a large warehouse at the University of Minnesota and assembled into 32-layer blocks at Ash River. Beneficial occupancy of the FD building was obtained in April 2011 and the power and network infrastructure is in place. Figure~\ref{fig:fardetView} is an aerial view of the completed far detector building. The first FD block is expected to be installed during January 2012.

\section{NO$\nu$A Physics Goals}
\begin{figure}[ht]
\centering
\includegraphics[width=80mm]{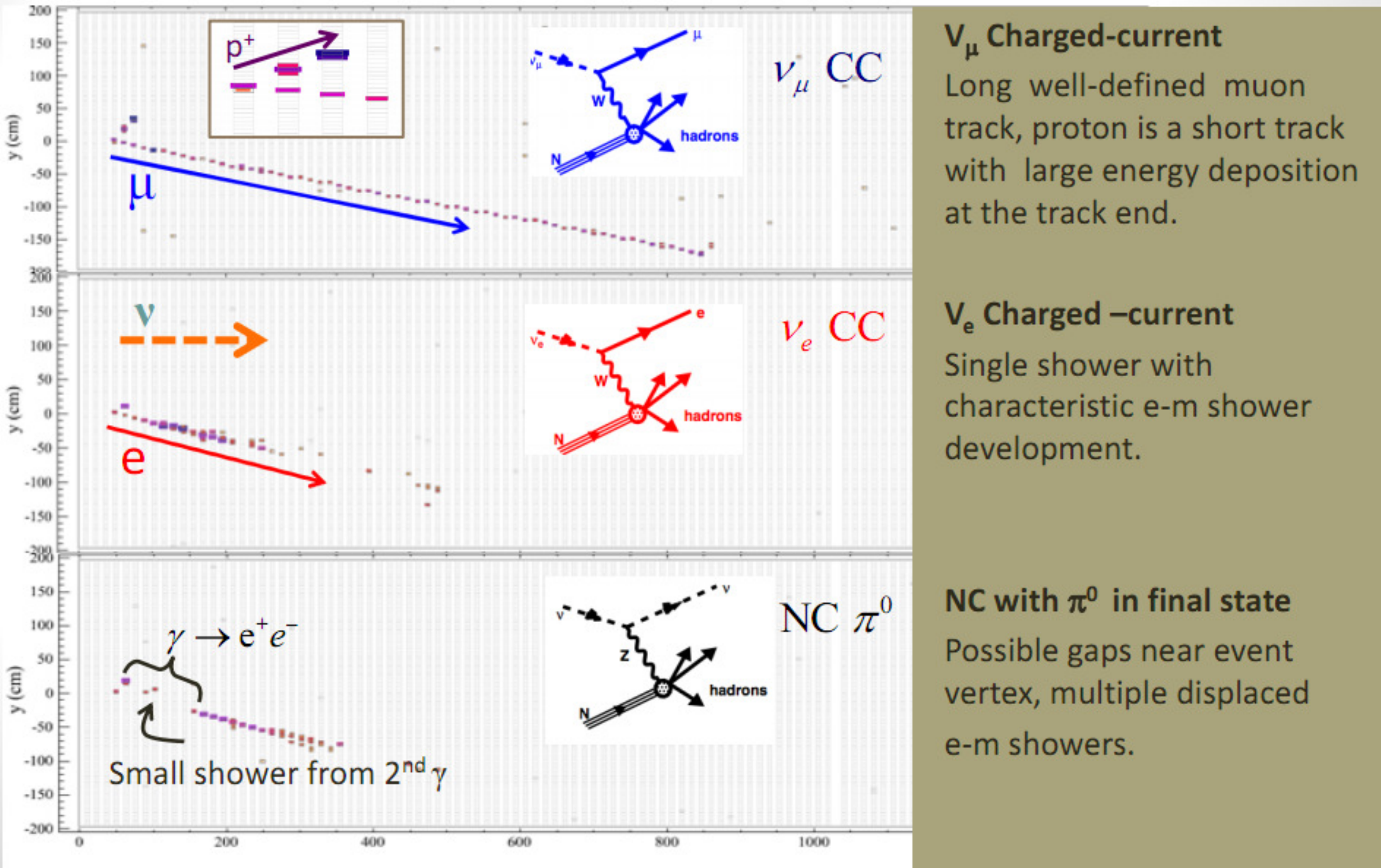}
\caption{Monte Carlo simulated event topologies in the NO$\nu$A detectors for $\nu_{\mu}$ CC, $\nu_{e}$ CC and NC with $\pi^{0}$ production neutrino interactions.} \label{fig:mc_events}
\end{figure}

NO$\nu$A will be able to distinguish $\nu_{e}$-CC interactions from backgrounds very effectively due to the low-density and high granularity of its detectors. Topologies for different categories of events simulated in the NO$\nu$A detectors are shown in Fig.~\ref{fig:mc_events}. The main backgrounds are expected to originate from NC events where a $\pi^{0}$ is produced and from the intrinsic beam $\nu_{e}$ component. The $\pi^{0}$ background events can be identified if spatial gaps are apparent near the event vertex and/or the event contains multiple displaced electromagnetic showers created by the two photons the $\pi^{0}$ decays into. The intrinsic beam $\nu_{e}$ component can be estimated from the beam simulation and ND measurements.

\begin{figure}[ht]
\centering
\includegraphics[width=80mm]{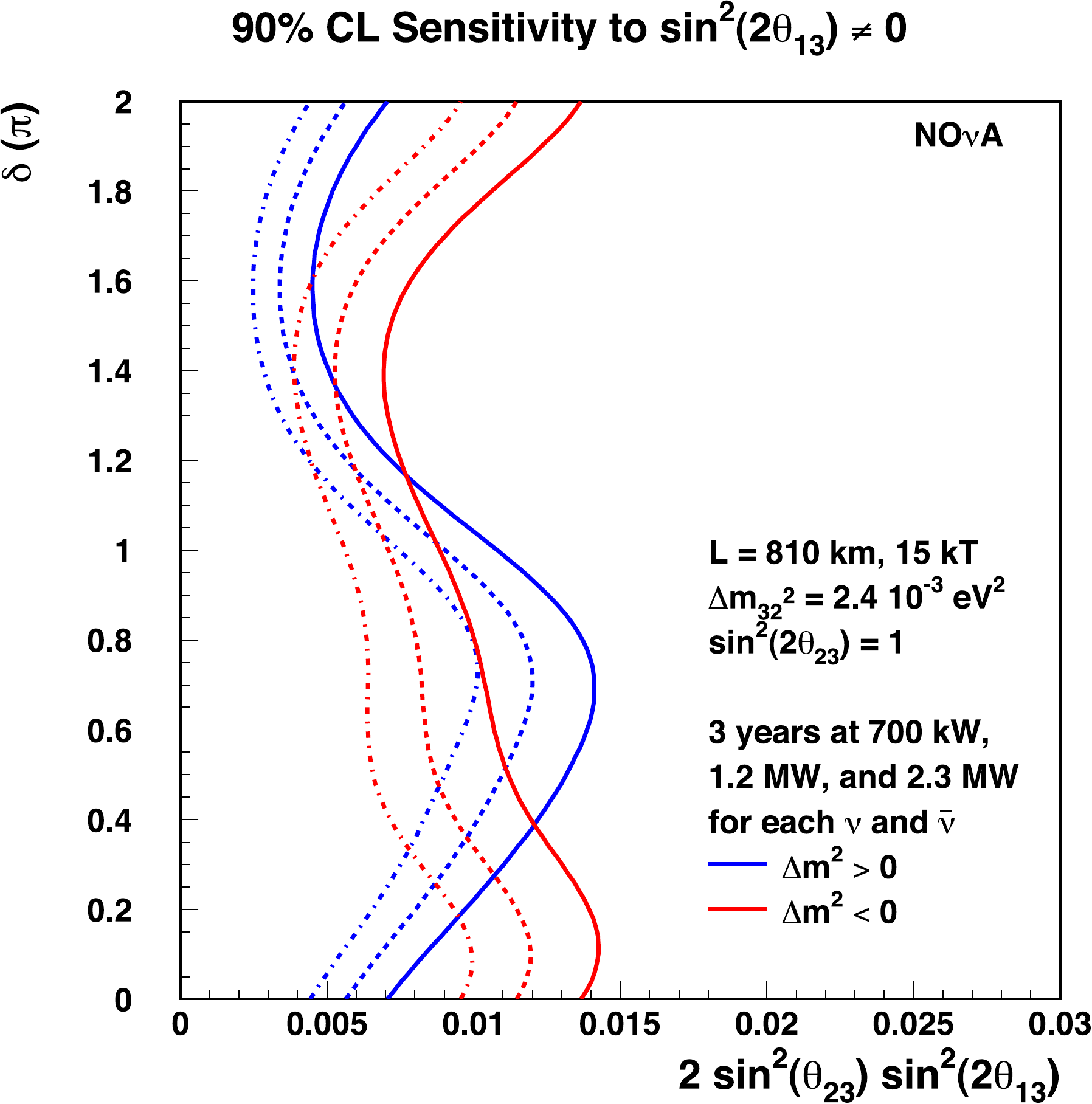}
\caption{NO$\nu$A's sensitivity to $\nu_{e}$ appearance as a function of $\delta_{CP}$. The blue curves assume normal mass hierarchy while the red curves show the inverted hierarchy case. The sensitivity is calculated assuming a 15~kT detector, 10\% systematic error on the backgrounds, and 6 years of running split evenly between neutrino and antineutrino horn polarities.} \label{fig:sens13}
\end{figure}

NO$\nu$A is sensitive to electron neutrino appearance down by an order of magnitude of the limit set by the CHOOZ \cite{chooz} experiment at 90\% CL. Figure~\ref{fig:sens13} shows NO$\nu$A's sensitivity to $\sin$(2$\theta_{13}$) after 3 years each of running neutrino and antineutrino beams. Contours for different beam upgrades are also shown with an assumed 18 x 10$^{20}$ POT of data in each neutrino and antineutrino mode. The recent T2K \cite{t2k13} and MINOS \cite{minos13} results are encouraging for NO$\nu$A as they both reported evidence for non-zero $\theta_{13}$. NOvA is also sensitive to an order of magnitude better than the recent global fit which suggests that $\theta_{13}$ $>$ 0 \cite{fogli}.

\begin{figure}[ht]
\begin{minipage}[b]{0.5\linewidth}
\centering
\includegraphics[width=80mm]{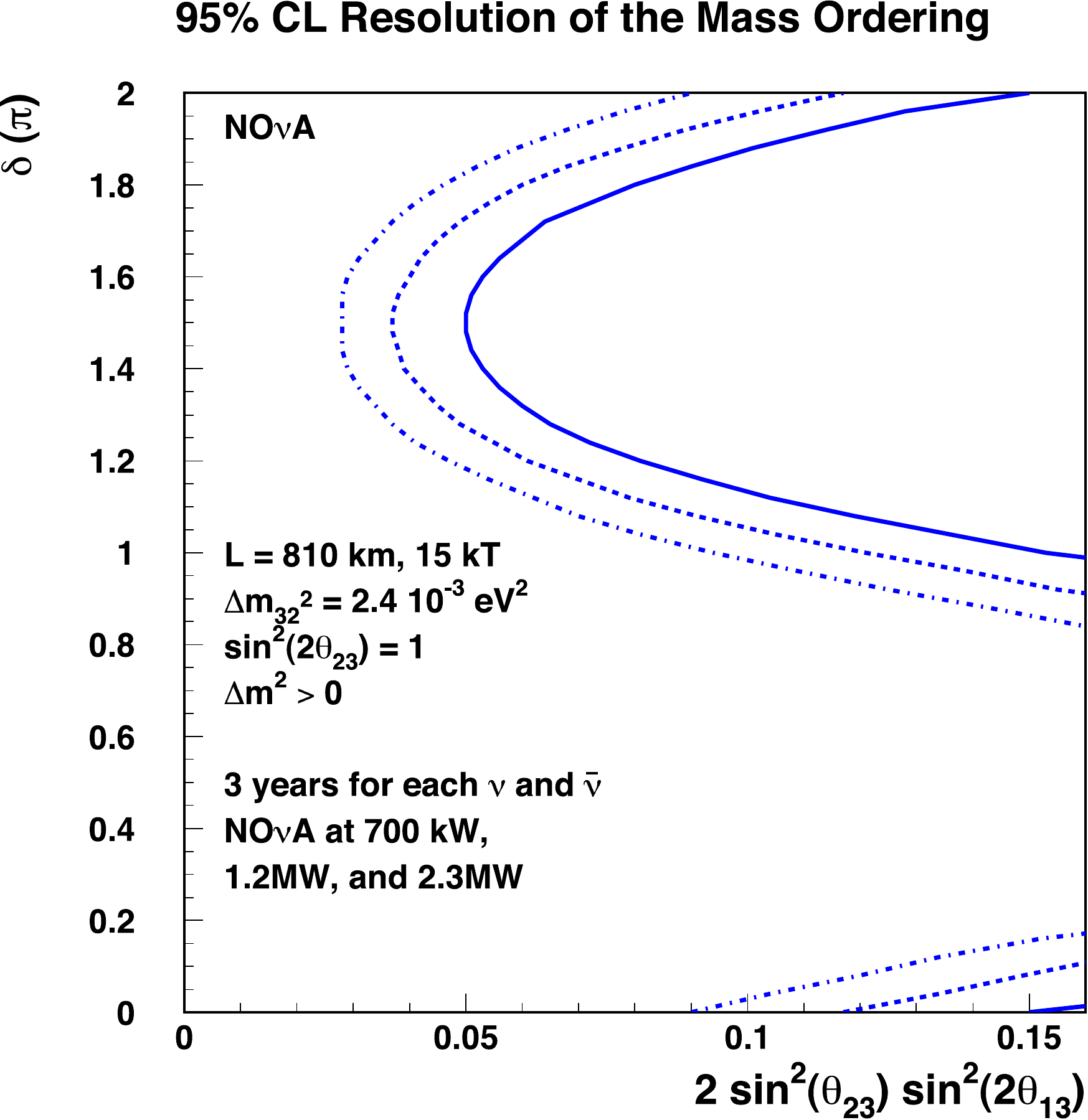}
\end{minipage}%
\begin{minipage}[b]{0.5\linewidth}
\centering
\includegraphics[width=80mm]{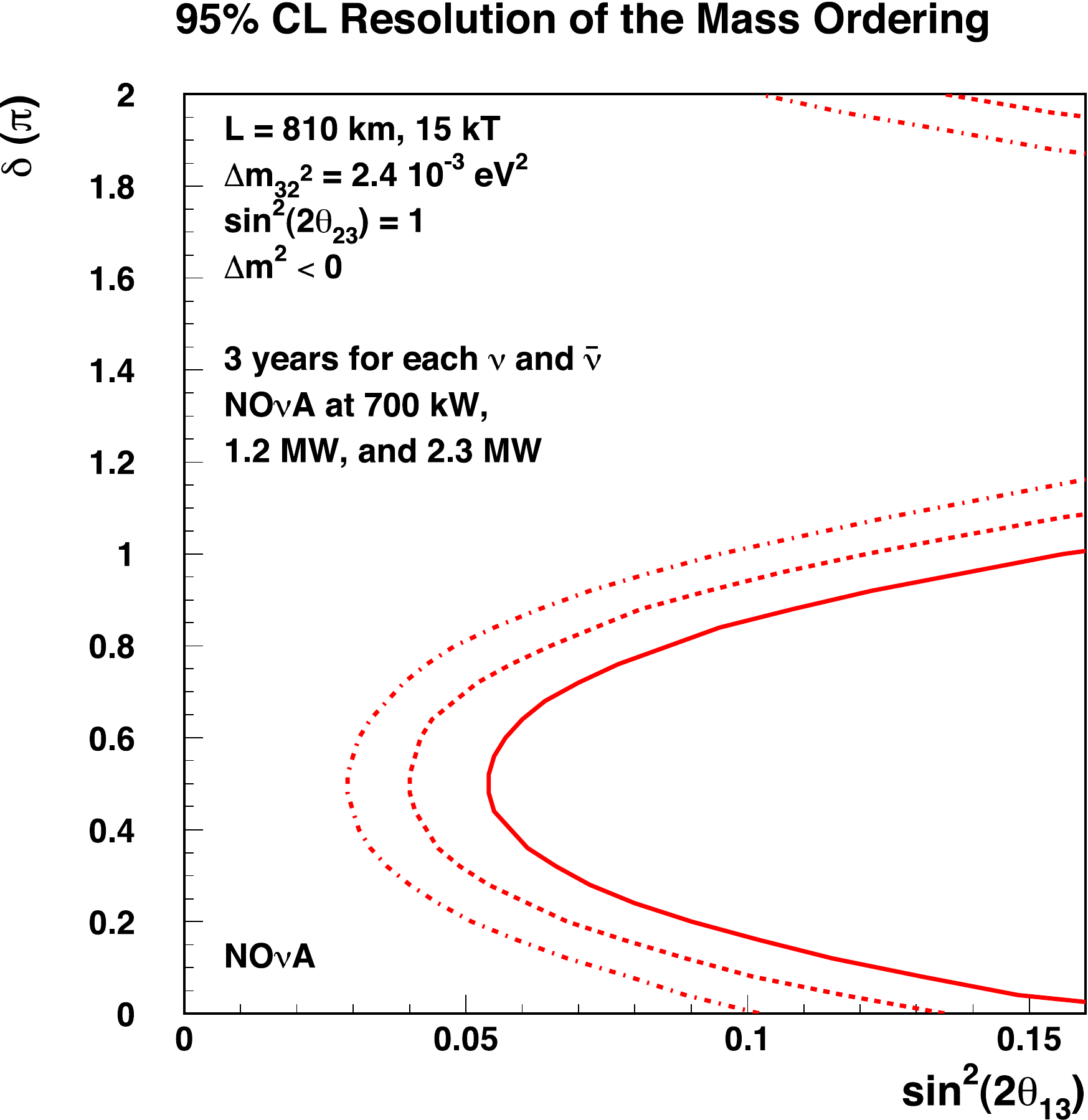}
\end{minipage}
\caption{For oscillation parameters to right of these curves, NOvA resolves the neutrino mass hierarchy with better then 95\% C.L. The curves are calculated for a 15~kT detector, 6 years of running split evenly between neutrino and anti-neutrino horn polarities. Intensities of the baseline 700~kW and possible further upgrades to 1.2~MW and 2.3~MW are also shown. The plot assumes nature has a normal mass hierarchy (left) or inverted mass hierarchy (right).}
\label{fig:massh}
\end{figure}

Due to NO$\nu$A's 810~km baseline, matter-induced oscillations affect the neutrino oscillation probability with a 30\% enhancement or suppression depending on the value of $\theta_{13}$ and $\delta_{CP}$. The matter effects also depend on the mass hierarchy, described by the sign of $\Delta m^2_{31}$, and change P($\nu_{\mu} \rightarrow \nu_{e}$) and P($\overline{\nu}_{\mu} \rightarrow \overline{\nu}_{e}$) differently. By running for 3 + 3 years ($\nu$ and $\overline{\nu}$) NO$\nu$A may resolve the mass hierarchy if $\theta_{13}$ is large enough. Figure~\ref{fig:massh} shows NO$\nu$A's resolution of the normal (left) and inverted (right) neutrino mass hierarchy respectively at the 95\% C.L.
 
\begin{figure}[ht]
\centering
\includegraphics[width=135mm]{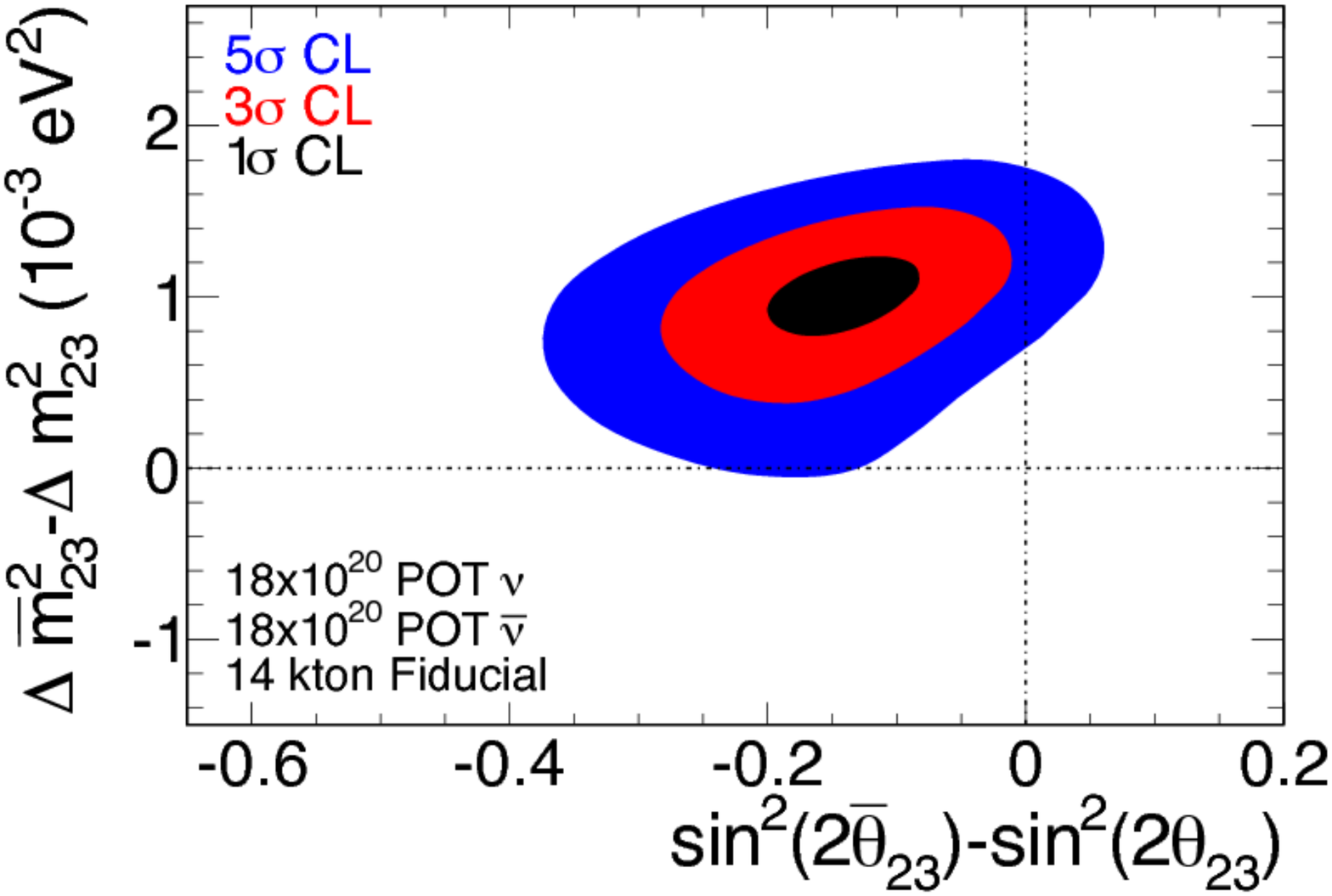}
\caption{NO$\nu$A contours for the difference in neutrino and anti-neutrino muon neutrino disappearance parameters. The difference is set to coincide with the recent difference reported by MINOS at 2$\sigma$ C.L \cite{minos23}. Contours are based on three years of neutrino running and three years of anti-neutrino running and use quasi-elastic events.} \label{fig:contours}
\end{figure}

NO$\nu$A will also be able to signiﬁcantly improve the precision on the measurements of atmospheric neutrino oscillation parameters $\Delta m^2_{32}$ and $\theta_{23}$. MINOS reported a $\sim$2$\sigma$ difference between best fit values for $\nu$ and $\overline{\nu}$ disappearance parameters in their 2010 $\overline{\nu}_{\mu}$ analysis \cite{minos23}. NO$\nu$A would be able to confirm (at more than 3$\sigma$) or rule out the asymmetry between $\Delta m^2_{32}$ and $\Delta \overline{m^2_{32}}$ hinted at by MINOS with 1 year of neutrino running and 1 year of antineutrino running. NO$\nu$A could establish this difference at the 5$\sigma$ level after the full 6 year run, 3 + 3 years in neutrinos and antineutrinos. This is reflected in Fig.~\ref{fig:contours}. It should be noted that MINOS recently released new antineutrino oscillations results \cite{minosNEW} whereby they include an increased antineutrino data exposure. This new MINOS 2011 analysis shows that neutrinos and antineutrinos are now consistent at the 42\% C.L. The previous $\sim$2$\sigma$ difference appears to have been a fluctuation. NO$\nu$A will update its expected disapperance parameter measurements using the latest MINOS best fit values in due course.

\section{NDOS Status}
The Near Detector On the Surface (NDOS) is the prototype Near Detector. It is functionally identical to the ND and has been operating on the surface at Fermilab and taking neutrino data since October 2010. The NDOS location at Fermilab has been chosen so that it is simultaneously exposed to the NuMI neutrino beam (110~mrad or 6.4\textdegree off-axis) and the Booster neutrino beam ($\sim$ on-axis). The NDOS was built to mimic far site construction as closely as possible. Building the NDOS allowed the factories and collaboration to fully exercise their quality assurance/quality control (QA/QC) techniques in preparation for full production running for the far detector. This process revealed that $\sim$22\% of the manifold covers cracked after assembly into a block. These covers have been repaired and a new more robust design with more rigorous control has been adopted for future production. Furthermore, oil work allowed us to gain experience qualifying the scintillating oil and filling the detector modules with the oil. Surface cleanliness and sealing issues have led to many of the NDOS APDs becoming unusably noisy. 274 installed APD units have been removed from the NDOS detector for cleaning and study. The collaboration is investigating new surface coating and installation techniques which will alleviate these issues prior to far detector
construction.

\subsection{Results from NDOS}
At peak performance, the NDOS ran warm with 75\% of its available channels readout. At the time of writing only 45\% of the detector is live. Despite this, the NDOS provides extremely valuable preparation for construction at Ash River. It also has allowed us to have an early look at real cosmic rays and neutrinos and ultimately given us a headstart on calibration techniques and physics analyses.

\begin{figure}[ht]
\centering
\includegraphics[width=135mm]{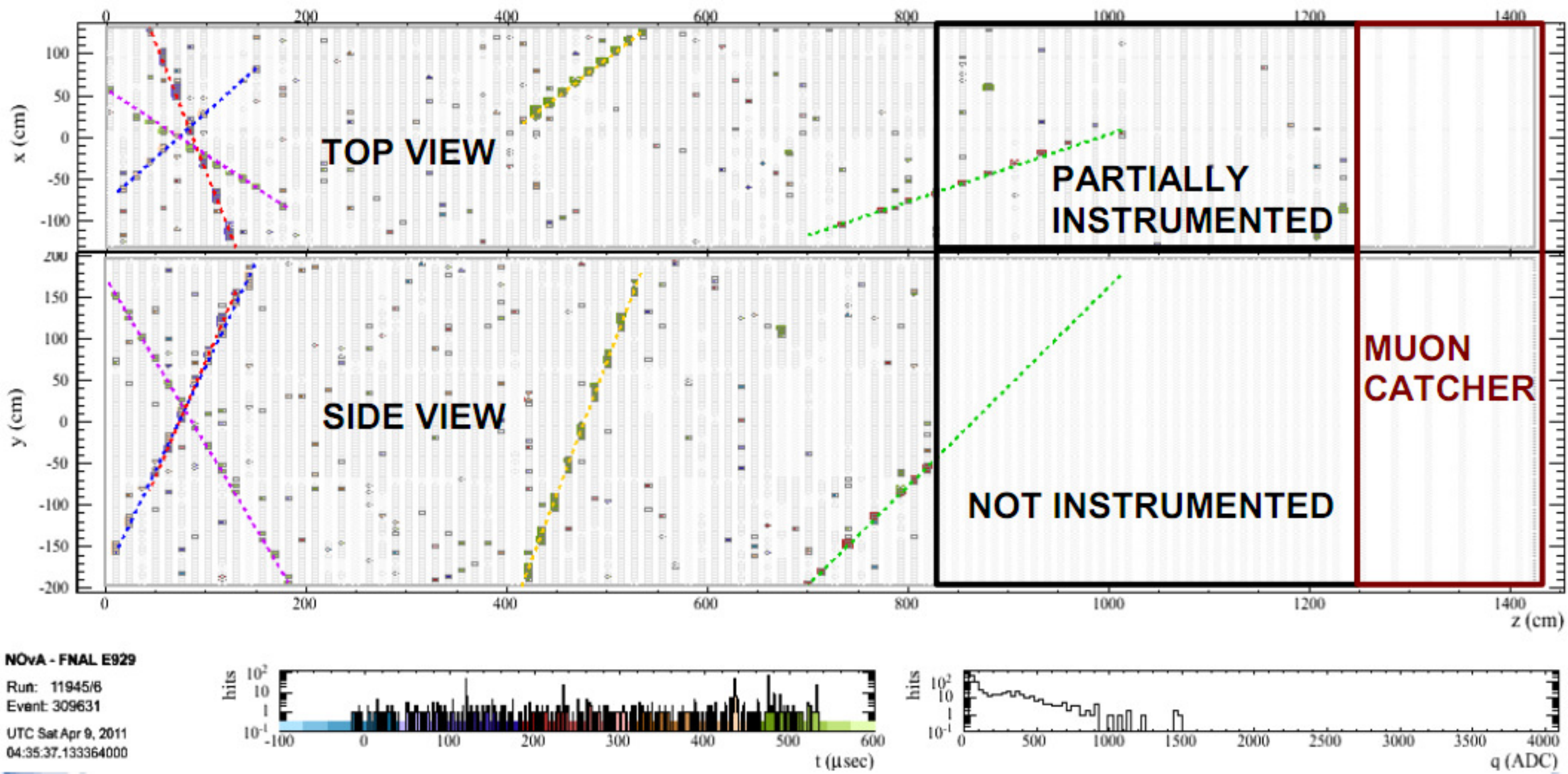}
\caption{Event Display showing cosmic data taken with the NDOS detector. The Event Display shows the
top and side views of the detector representing vertical and horizontal modules. Sections that are not instrumented are shown in light gray.} \label{fig:cosmic}
\end{figure}

\begin{figure}[ht]
\begin{minipage}[b]{0.5\linewidth}
\centering
\includegraphics[width=80mm]{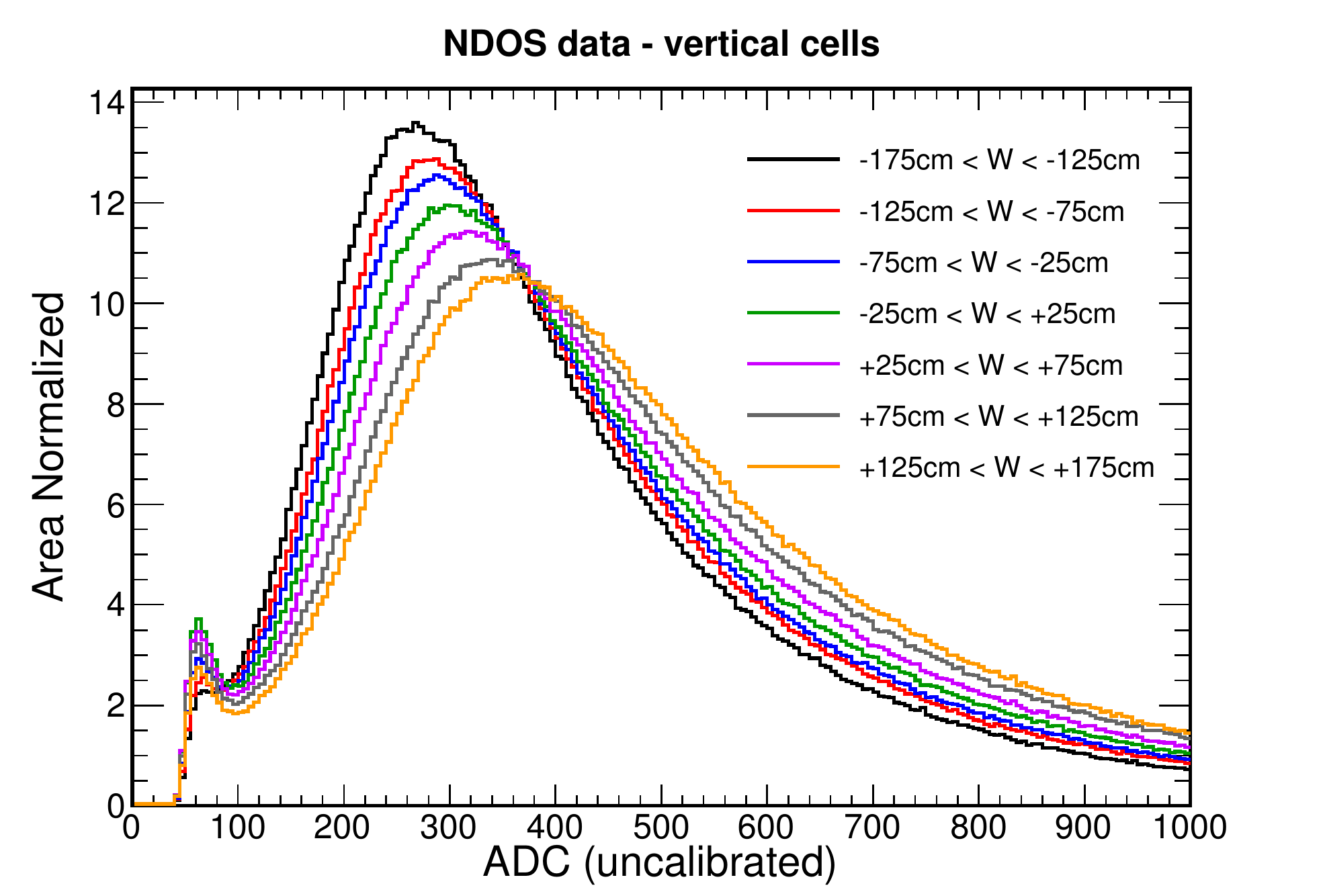}
\end{minipage}%
\begin{minipage}[b]{0.5\linewidth}
\centering
\includegraphics[width=80mm]{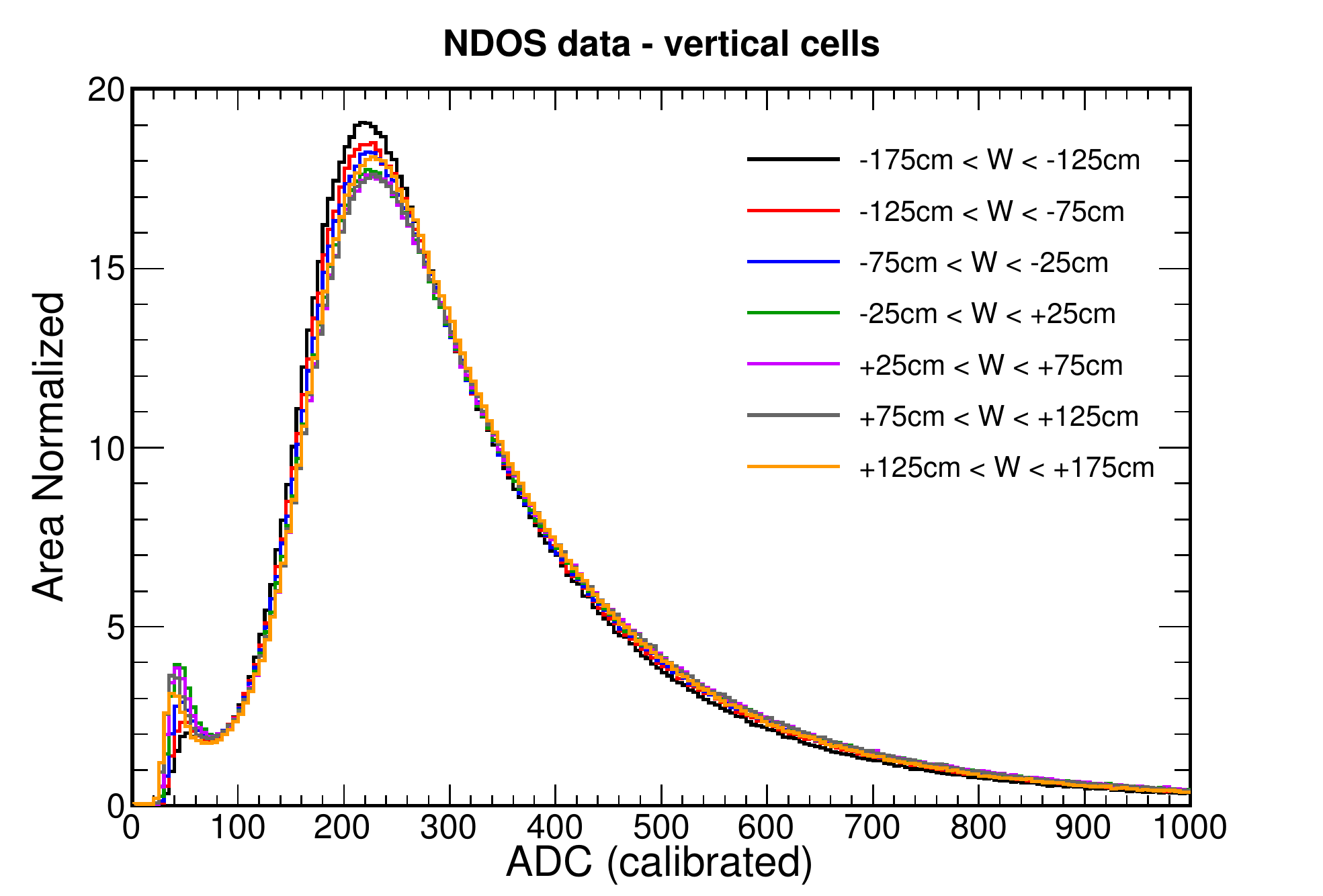}
\end{minipage}
\caption{Path length-corrected muon response for different distances from wavelength-shifting fiber end for a single example cell (left). Muon response after attenuation corrections (right).}
\label{fig:adc}
\end{figure}

Figure~\ref{fig:cosmic} is an example event display of cosmic ray data taken with the NDOS detector. The color corresponds to hits belonging to the same track and the line indicates the fit result of the cosmic ray reconstruction algorithm. The reconstructed cosmic rays are used for commissioning, detector alignment, calibration and stability studies. The cosmic ray reconstructed tracks have been utilised to calibrate corrections for the differences in signal
pulse heights from tracks going through cells at different distances from the APD sensors caused by fiber attenuation. Figure~\ref{fig:adc} shows the result of measuring the path-length corrected muon response for an example cell and then the effect of applying calculated attenuation corrections to produce a calibrated muon response for the cell.

\begin{figure}[ht]
\centering
\includegraphics[width=135mm]{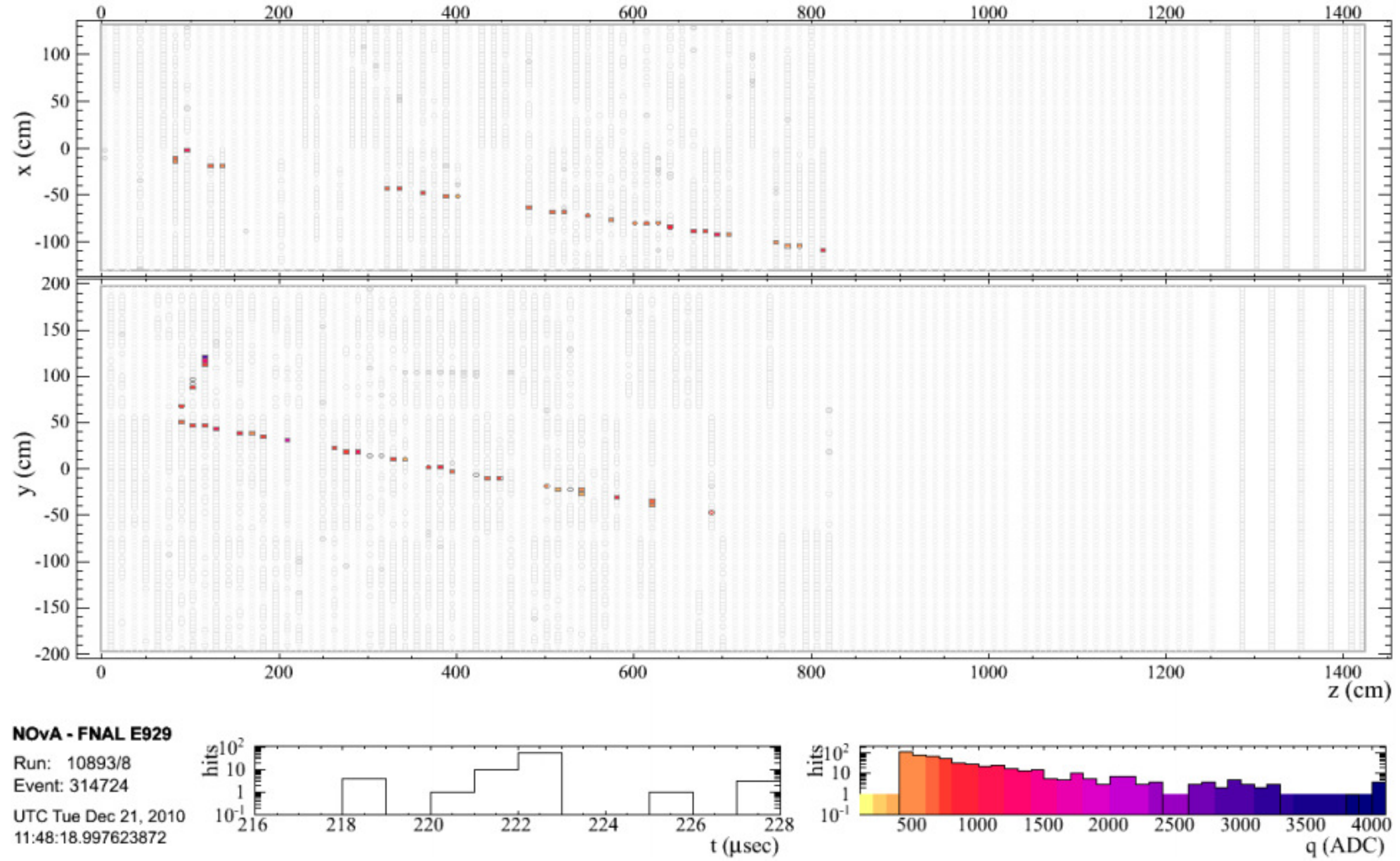}
\caption{Event Display showing example candidate neutrino event from the NuMI source in NDOS.} \label{fig:neutrinoEVD}
\end{figure}

An example neutrino event for NuMI is shown in Fig.~\ref{fig:neutrinoEVD}. This is a candidate $\nu_{\mu}$ charged-current event which can be seen when comparing to Fig.~\ref{fig:mc_events}. One can clearly see the long well-defined muon track in both detector views as well as the smaller hadronic track that leaves a large energy deposition at its track end. All hits in these tracks occur within a 3~$\mu$s window. NDOS has collected 5.6 x 10$^{19}$ POT worth of data in reverse horn current beam and 8.4 x 10$^{18}$ POT in forward horn current mode from NuMI. A subsequent analysis of this sample has yielded 1001 and 253 candidate neutrino events with 69 and 39 expected cosmic background events respectively.

\section{Conclusion}
NO$\nu$A is on track to make many important contributions to neutrino physics such as the measurement of $\theta_{13}$, determination of the neutrino mass hierarchy and more precise measurements of $\Delta m^2_{32}$ and $\sin^{2}$(2$\theta_{23}$). Recent results from the T2K and MINOS experiments are very encouraging for the NO$\nu$A goals as they provide evidence for a non-zero $\theta_{13}$. The Far detector laboratory is complete and the detector is expected to start data-taking in 2012 as the construction progresses due to its modular nature. Construction of the full 14~kt is expected to be completed by early 2014. The NuMI beam upgrades are on schedule to begin during the accelerator shutdown in March 2012 when the ND cavern excavation will occur also. The NO$\nu$A NDOS has been taking data and analysing data since October 2010 and continues to operate and provide critical feedback for design enhancements and operational experience.


\bigskip 

\end{document}